\documentclass[letterpaper, 10 pt, journal, twoside]{IEEEtran}
\usepackage{amsmath,amsfonts}
\usepackage{algorithmic}
\usepackage{algorithm}
\usepackage{array}
\usepackage[caption=false,font=normalsize,labelfont=sf,textfont=sf]{subfig}
\usepackage{textcomp}
\usepackage{stfloats}
\usepackage{url}
\usepackage{graphicx, multirow, booktabs, ctable, pifont}
\usepackage{verbatim}
\usepackage{cite}
\usepackage{hyperref}
\usepackage{balance}
\usepackage[symbol]{footmisc}
\newcommand{\cmark}{\ding{51}}

\newcommand{\changeVariable}{}

\begin{document}

\title{Global-Reasoned Multi-Task Learning Model for Surgical Scene Understanding}

% Make room for more info lines in the \author command 
\author{Lalithkumar Seenivasan$^{1+}$, Sai Mitheran$^{2+}$, Mobarakol Islam$^{3}$, Hongliang Ren$^{1,4*} $ \textit{Senior Member, IEEE}
\thanks{Manuscript received: September, 9, 2021; Revised December, 12, 2021; Accepted January, 18, 2022.}%Use only for final RAL version
\thanks{This paper was recommended for publication by Editor Jessica Burgner-Kahrs upon evaluation of the Associate Editor and Reviewers' comments.}

\thanks{This work was supported by the National Key R\&D Program of China under Grant 2018YFB1307700 (with subprogram 2018YFB1307703) from the Ministry of Science and Technology (MOST) of China, Hong Kong Research Grants Council (RGC) Collaborative Research Fund (CRF C4026-21GF), the Shun Hing Institute of Advanced Engineering (SHIAE project BME-p1-21, 8115064) at the Chinese University of Hong Kong (CUHK), and Singapore Academic Research Fund under Grant R397000353114.} %Use only for final RAL version

\thanks{$^{+}$Lalithkumar Seenivasan and Sai Mitheran are co-first authors.}%
\thanks{$^{*}$Corresponding author.}%: {\tt\footnotesize hlren@ieee.org}}
\thanks{$^{1}$Lalithkumar Seenivasan and Hongliang Ren are with Dept. of Biomedical Engineering, National University of Singapore, Singapore. (\href{mailto:lalithkumar_s@u.nus.edu}{lalithkumar\_s@u.nus.edu}, \href{mailto:hlren@ieee.org}{hlren@ieee.org})}
\thanks{$^{2}$Sai Mitheran is with Dept. of Electronics and Communication Engineering, National Institute of Technology, Tiruchirappalli, India.
(\href{mailto:108118084@nitt.edu}{108118084@nitt.edu})}
\thanks{$^{3}$Mobarakol Islam is with Biomedical Image Analysis Group, Imperial College London, UK. (\href{mailto:mi615@ic.ac.uk}{mi615@ic.ac.uk})}
\thanks{$^{4}$Hongliang Ren is with Dept. of Electronic Engineering, The Chinese University of Hong Kong. (\href{mailto:hlren@ee.cuhk.edu.hk}{hlren@ee.cuhk.edu.hk})}
\thanks{Digital Object Identifier (DOI): see top of this page.}
}
% Use only for final RAL version.

% Paper headers
\markboth{IEEE Robotics and Automation Letters. Preprint Version. Accepted January, 2022}
{Seenivasan \MakeLowercase{\textit{et al.}}: Global-Reasoned Multi-Task Learning Model for Surgical Scene Understanding} 
% Use only for final RAL version

% \IEEEpubid{0000--0000/00\$00.00~\copyright~2021 IEEE}
% Remember, if you use this you must call \IEEEpubidadjcol in the second
% column for its text to clear the IEEEpubid mark.

\maketitle

\begin{abstract}
Global and local relational reasoning enable scene understanding models to perform human-like scene analysis and understanding. Scene understanding enables better semantic segmentation and object-to-object interaction detection. In the medical domain, a robust surgical scene understanding model allows the automation of surgical skill evaluation, real-time monitoring of surgeon’s performance and post-surgical analysis. This paper introduces a globally-reasoned multi-task surgical scene understanding model capable of performing instrument segmentation and tool-tissue interaction detection. Here, we incorporate global relational reasoning in the latent interaction space and introduce multi-scale local (neighborhood) reasoning in the coordinate space to improve segmentation. Utilizing the multi-task model setup, the performance of the visual-semantic graph attention network in interaction detection is further enhanced through global reasoning. The global interaction space features from the segmentation module are introduced into the graph network, allowing it to detect interactions based on both node-to-node and global interaction reasoning. Our model reduces the computation cost compared to running two independent single-task models by sharing common modules, which is indispensable for practical applications. Using a sequential optimization technique, the proposed multi-task model outperforms other state-of-the-art single-task models on the MICCAI endoscopic vision challenge 2018 dataset. Additionally, we also observe the performance of the multi-task model when trained using the knowledge distillation technique. The official code implementation is made available in GitHub\footnote[3]{\href{https://github.com/lalithjets/Global-reasoned-multi-task-model}{github.com/lalithjets/Global-reasoned-multi-task-model}}\label{code}.
\end{abstract}

% Keywords appear just beneath the abstract. Use only for final RAL version. 

\section{Introduction}

\begin{figure}[!b]
\centering
\includegraphics[width=1.0\linewidth]{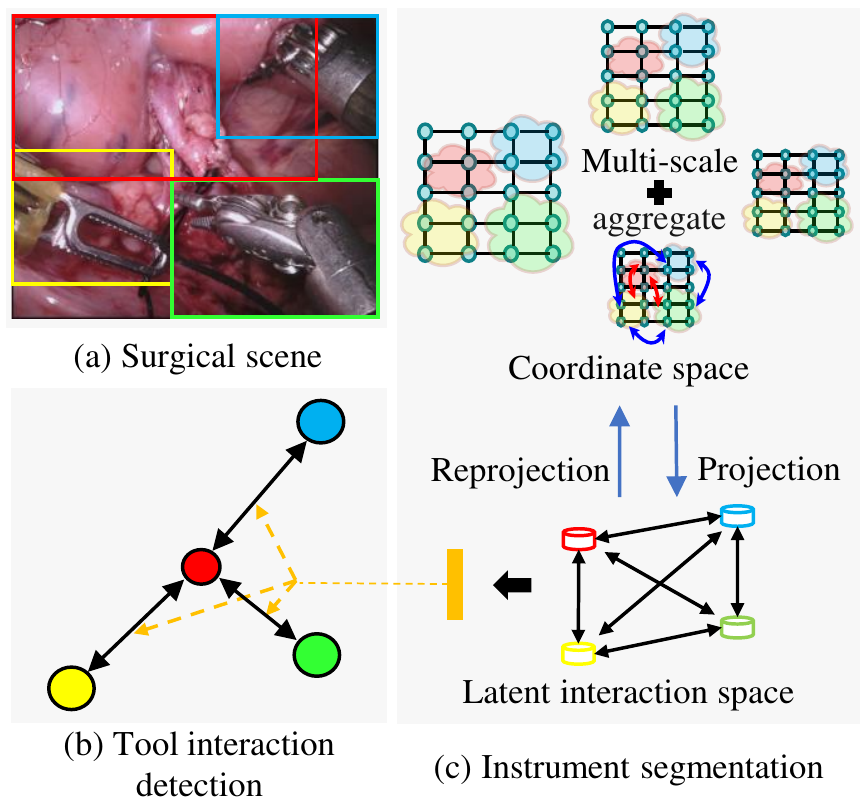}
\caption{Enhancing surgical scene understanding (tool interaction detection and instrument segmentation) through global-local relational reasoning.}
\label{fig:global_interaction_space}
\end{figure}

\IEEEPARstart{M}{imicking} a human’s ability in understanding and analyzing a situation requires scene understanding models to reason on both local and global relationships. Classification and regression models that work on images have immensely benefited from local relationship reasoning~\cite{10.5555/1005332.1016789}. In addition to local relationships, global scene awareness plays a crucial role in detecting object-to-object interaction~\cite{wang2017onlocal}, semantic segmentation~\cite{zhao2016yramid}, video segmentation, and video analysis. For instance, the action of a tool in a surgical scene could be detected as being ‘idle’ when not in contact with tissue. However, it could also be detected as ‘tool manipulation’ despite not being in contact with the tissue if used to hold another tool. Embedding the scene awareness to object detection and semantic segmentation models could result in the model awarding a higher likelihood to certain instrument classes depending on the scene. Extracting relational features in an image has been a key element of deep neural networks. While convolution (conv) layers have dominated other modules in capturing neighborhood (local) relations, capturing global relations often requires deeper networks, making the network computationally inefficient. To efficiently capture global relations, a graph-based global reasoning network~\cite{chen2019graph} was proposed that performs global reasoning in the latent space. Inspired by this work, we propose a global-reasoned multi-task surgical scene understanding model that performs instrument segmentation and detects tool-tissue interaction.

Global-reasoned surgical scene understanding is critical in developing surgical skill assessment, real-time and post-surgical analysis, augmented tactile feedback and automated surgical report generation. In the proposed model (Fig. \ref{fig:global_interaction_space}), we employ the global reasoning (GR) unit (GloRe unit)~\cite{chen2019graph} to improve instrument segmentation by reasoning the global relations in the latent interaction space. Since a GloRe unit performs reasoning in the latent space and cannot fully replace a fully connected graph in coordinate space, we further introduce multi-scale decoder aggregation for semantic segmentation. By combining the GloRe unit that reasons in the latent space and multi-scale-feature decoder aggregation that captures local relations at multiple scales, the semantic segmentation model is aimed to perform better scene reasoning. To detect the tool action, we improve upon the visual-semantic graph attention network (VS-GAT)~\cite{liang2020isualsemantic} and introduce Globally-Reasoned VS-GAT. While VS-GAT aims to detect node interaction through node-to-node reasoning, it still lacks global reasoning as its nodes are embedded only with features of tools or defective tissue. By embedding global-reasoned latent features to VS-GAT, we hypothesize the model to detect globally-reasoned node-to-node interaction.
Furthermore, taking advantage of the multi-task model, (i) the feature extractor trained for semantic segmentation is also employed to embed VS-GAT’s nodes and (ii) embed its edges with global-reasoned interaction space features (GISF) obtained from the segmentation model’s GloRe unit, improving model generalization. By sharing the feature encoder and GloRe unit, we also reduce the computational cost compared to running two independent single-task models. Our key contributions are as follows:
\begin{itemize}
\item Propose a globally-reasoned multi-task learning (MTL) surgical scene understanding model that performs instrument segmentation and tool-tissue interaction detection. 
    \item Improve the MTL model’s segmentation performance by incorporating latent global interaction reasoning and introducing multi-scale local reasoning.
    \item Utilize the MTL model setup to enhance interaction detection performance by sharing a generalized feature extractor for visual feature extraction and incorporating globally-reasoned features from the segmentation module into the scene graph (tool interaction detection) model.
   \item Study the performance of sequential and knowledge distillation (KD) based optimization techniques in optimizing MTL models for optimal model convergence.
\end{itemize}

\section{Related Work}
\label{RELATED WORKS}

% \subsubsection{Global reasoning}

\subsubsection{Surgical Instrument Segmentation}

The surgical instrument segmentation task has been predominantly addressed using conv backbones~\cite{chaurasia2017inknet, garcia-peraza-herrera2017oolnet, islam2019earning, 9196905, shvets2018automatic, ronneberger2015net}. These methods employ an encoder-decoder architecture to perform pixel-wise segmentation. In performing pixel-wise segmentation, they suffer from spatial inconsistency where pixels of a single instrument could be assigned to other instruments. To address this, instance-based segmentation has also been proposed~\cite{8877379, jin2019ncorporating}. Current state-of-the-art (SOTA) models in instrument segmentation include MF-TAPNet~\cite{jin2019ncorporating} and ISI-Net~\cite{gonzalez2020sinet}. MF-TAPNet~\cite{jin2019ncorporating} employs an attention mechanism and utilizes temporal optical flow. Built on top of Mask-RCNN~\cite{he2017ask}, ISI-Net~\cite{gonzalez2020sinet} employs a temporal consistency strategy to take advantage of the temporal frame sequence. A refined attention-based network called RASNet~\cite{Ni2019RASNetSF} was also proposed that utilizes the attention mechanism for semantic segmentation to leverage on the global context of high-level features to focus on key regions of the image. It employed a novel module to fuse low-level features with the global context from high-level features. As an alternative to these prior works, we propose a simple and efficient global and local reasoned model that achieves competitive performances against existing SOTA models.

\subsubsection{Surgical Tool Interaction Detection}
Object-to-object interaction detection has been a key focus in scene understanding. Initially, human-to-object interaction detection was achieved by employing Fast-RCNN~\cite{Mallya2016LearningMF} and Faster-RCNN~\cite{Gkioxari2018DetectingAR}. However, these methods are not robust to dynamic scenes where the number of objects is not constant. This issue was addressed by theorizing the interaction detection task in the non-euclidean space and employing graph networks to detect interaction~\cite{Qi2018LearningHI, liang2020isualsemantic}. Graph passing neural network (GPNN)~\cite{Qi2018LearningHI} theorizes each scene as a sparse graph, with its nodes being the objects and edges denoting the presence of interaction. In the medical domain, Islam et-al~\cite{islam2019earning} improved GPNN by integrating with GraphSAGE~\cite{hamilton2017nductive} and label smoothing~\cite{muller2019hen} for tool-tissue interaction detection. While GPNN relies mainly on visual features to detect object-to-object interaction, the visual semantic graph attention network (VS-GAT)~\cite{liang2020isualsemantic} was introduced that utilizes spatial and semantic features on top of visual features to detect interactions. Here, we further improve the VS-GAT in detecting interaction by including globally-reasoned features.

\subsubsection{Multi-Task learning}
A single MTL model offers a computational advantage over multiple single-task learning (STL) models. However, during training, the MTL model is affected by asynchronous convergence in its independent task performance. To address this, various MTL optimization techniques have been proposed. GradNorm~\cite{chen2017radnorm} helps balance the learning of independent task sub-modules, thereby balancing independent task influence on the shared module and improving synchronization in model convergence. Attention prone MTL optimization technique~\cite{islam2019earning} has also been proposed that enables sequential convergence of model’s independent tasks. Recently, KD-based MTL optimization has also been proposed~\cite{li2020knowledge}. In this work, we also implement and study the performance of (i) sequential, (ii) vanilla, and (iii) KD-based MTL optimization in training our proposed MTL model.

\begin{figure*}[!h]
\centering
\includegraphics[width=1.0\linewidth]{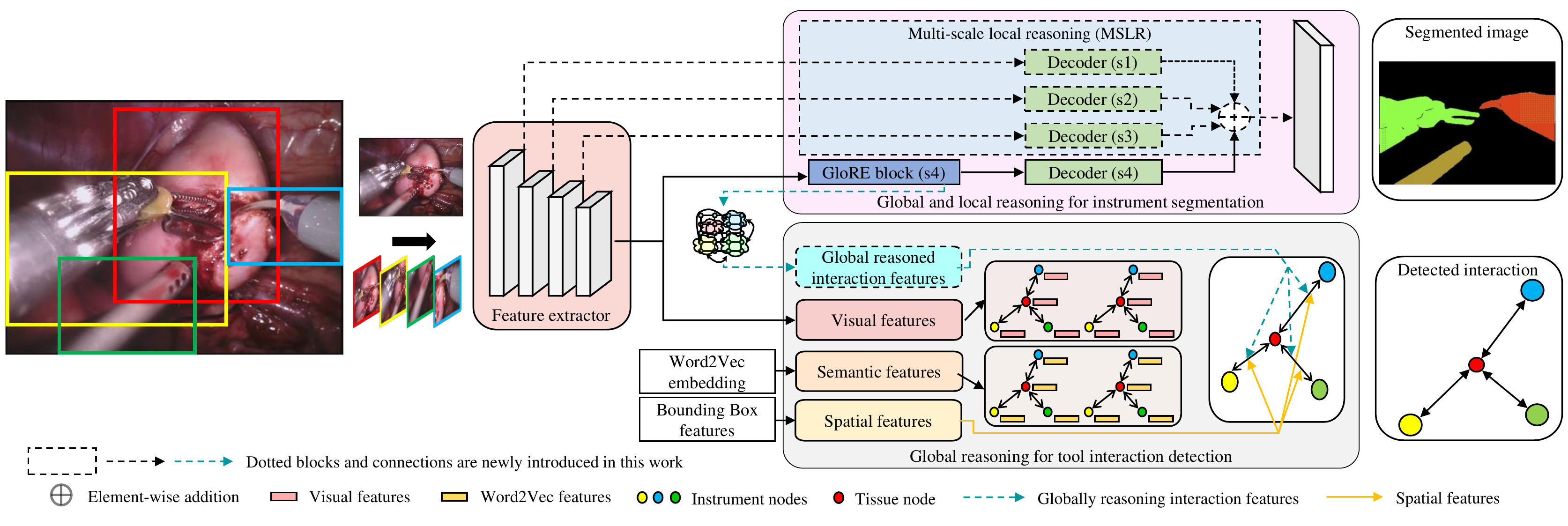}
\caption{The proposed network architecture. The proposed globally-reasoned multi-task scene understanding model consists of a shared feature extractor. The segmentation module performs latent global reasoning (GloRe~\cite{chen2019graph} unit) and local reasoning (multi-scale local reasoning) to segment instruments. To detect tool interaction, the scene graph (tool interaction detection) model incorporates the global interaction space features to further improve the performance of the visual-semantic graph attention network~\cite{liang2020isualsemantic}.}
\label{fig:architecture}
\end{figure*}

\section{Methodology}
\label{Proposed Solution}
We propose a globally-reasoned multi-task surgical scene understanding model to perform instrument segmentation and detect tool-tissue interactions. As shown in Fig.~\ref{fig:architecture}, the model consists of three main sub-modules: (a) Feature encoder, (b) Global and local reasoning for instrument segmentation and (c) Global reasoning for interaction detection. As the first sub-module, a ResNet18 model is employed to extract the features from the scene for segmentation and scene graph sub-modules. The segmentation sub-module consists of multi-scale feature connections between feature encoder, GloRe unit, and decoder modules. The interaction detection module consists of the globally-reasoned VS-GAT. The VS-GAT~\cite{liang2020isualsemantic} module utilizes the visual (extracted using the feature extractor), semantic and spatial features to detect interactions. This work further improves it by incorporating global interaction space features (GISF) in its graph aggregation.

\subsection{Global and Local Reasoning for Instrument Segmentation}
A simple encoder-decoder pair is employed to achieve competitive performance with the SOTA models in instrument segmentation. This performance is further enhanced by incorporating global reasoning. Here, global reasoning is achieved through a two-tier mechanism. Initially, the GloRe unit~\cite{chen2019graph} is employed to reason global interaction in the latent space. While this reasoning is limited to the latent interaction space, we include the Multi-Scale Local Reasoning module (MSLR). Here, multi-scale decoder aggregation is performed to capture multi-scale local (neighborhood) relations in coordinate space. For the decoder block, we design a lightweight decoder with (a) a conv block (conv-BatchNorm-ReLU), (b) a dropout and (c) finally a conv layer.

\begin{figure}[!b]
\centering
\includegraphics[width=0.75\linewidth]{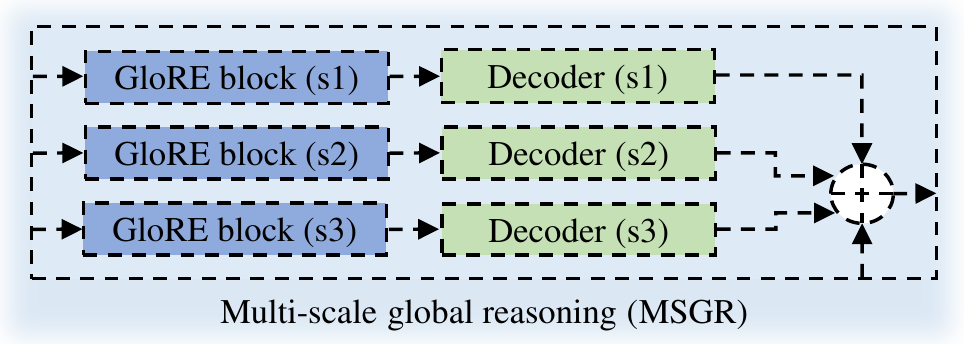}
\caption{Multi-scale global reasoning for instrument segmentation}
\label{fig:fig3}
\end{figure}

\subsubsection{Variants in Global Reasoning for Instrument Segmentation}
To improve instrument segmentation, three variants of global reasoning, (i) vanilla GR, (ii) Multi-scale global reasoning (MSGR), and (iii) multi-scale local reasoning and GR (MSLRGR), have been studied. Under vanilla GR, the GloRe unit~\cite{chen2019graph} is naively implemented to reason on the encoder's latent features. In MSGR, the GloRe unit is employed to reason on multi-scale interactions as shown in Fig.~\ref{fig:fig3}. The multi-scale reasoned features are then passed to their respective scale-specific decoder block. Finally, in MSLRGR (Fig. \ref{fig:architecture}), global reasoning is achieved by combining vanilla GR and multi-scale local (neighborhood) reasoning (MSLR). Considering the multi-scale encoder features locally reasoned through conv, these features are passed to their respective scale decoder block. These multi-scale decoder block outputs are then aggregated for pixel-wise segmentation.

\subsection{Global Reasoning for Interaction Detection}
The VS-GAT~\cite{liang2020isualsemantic} network employs two sub-graphs: (a) Visual graph ($\mathcal{G}_v$) and (b) Semantic graph ($\mathcal{G}_s$) embedded with visual ($\mathcal{F}_{vf}$) and semantic features ($\mathcal{F}_{semf}$), respectively. The two graphs are then propagated and fused to form a combined graph ($\mathcal{G}_c$). The edges of this graph are embedded with spatial features ($\mathcal{F}_{sf}$ = features of bounding box location). As the visual graph’s nodes are only embedded with tools and defective tissue features, it could be inferred as only node-node reasoning rather than global reasoning. Taking advantage of the multi-task model setup, we append the GISF ($\mathcal{F}_{GISF}$) from the segmentation module’s GloRe unit to the combined graph’s edges. This allows the model to predict the interactions based on both node-to-node and global latent interaction reasoning  [$\mathcal{Y} = \mathcal{G}(\mathcal{F}_{vf}, \mathcal{F}_{semf}, \mathcal{F}_{sf}, \mathcal{F}_{GISF})$].

\subsection{Multi-Task Optimization}

Optimizing a Multi-task model is challenging due to asynchronous convergence of its independent task performance. To address this, we explore three different optimization techniques. The first one, Vanilla-MTL (V-MTL) optimization, naively combines the loss of both tasks during the training (Eq. \ref{V_MTL_loss}). Here, by setting $\alpha = 0.4$, the losses of interaction detection ($\mathcal{L}_{sg}$) and segmentation ($\mathcal{L}_{seg}$) are scaled to 0.4 and 0.6, respectively, as the scene graph network is observed to converge faster in the single-task model setup.

\begin{equation}
\label{V_MTL_loss}
\mathcal{L}_{V-MTL} = ( \alpha * \mathcal{L}_{sg}) + ( (1 - \alpha) * \mathcal{L}_{seg})
\end{equation}

In the second variant, KD-based MTL (KD-MTL) optimization~\cite{li2020knowledge} is explored. The segmentation model’s reliance on the feature encoder for its performance was significantly higher, and its performance convergence was also slower in the STL model. To address this, the KD-MTL favors the segmentation task in training the feature encoder. Here, the task losses are combined with Kullback-Leibler divergence (KLD) ~\cite{10.1214/aoms/1177729694} loss between the feature encoder outputs of the STL segmentation model and MTL model. By reducing the KLD loss between the outputs of the feature encoder, we aim for the MTL to improve model convergence of the segmentation model.

\begin{equation}
\mathcal{L}_{KD-MTL} = (\alpha * \mathcal{L}_{sg}) + \mathcal{L}_{seg} + \mathcal{L}_{KLD-seg}
\end{equation}
where $\alpha = 0.4$ and $\mathcal{L}_{KLD-seg}$ is the KLD loss.

Inspired from asynchronous optimization~\cite{islam2019earning}, the final optimization technique involves optimizing the MTL model sequentially (S-MTL). As shown in Algorithm \ref{algortihm_ato}, the MTL model’s feature encoder and segmentation model is first trained based on the segmentation loss. Upon convergence, the weights of the feature encoder and segmentation blocks are frozen. The training of scene graph in detecting interactions is then performed until convergence. 
 
\begin{algorithm}[!h]
    \caption{\small{S-MTL Optimization}}
    \begin{algorithmic}[1]
        \label{algortihm_ato}
        \small
        
        \STATE \textbf{[Initialize model weights]}\\shared feature extractor (${W_{sh}}$), scene segmentation (${W_{seg}}$), scene graph (${W_{sg}}$) \\
        
        \STATE \textbf{[Set gradient accumulators to zero]}\\shared feature extractor (${dW_{sh}}$),  scene segmentation (${dW_{seg}}$), scene graph (${dW_{sg}}$) \\
        $\mathbf{dW_{sh}} \leftarrow 0,\; \mathbf{dW_{seg}} \leftarrow 0,\; \mathbf{dW_{sg}} \leftarrow 0$\\
        
        \STATE \textbf{[Optimize feature extractor and segmentation network]}\\
        $\mathbf{while}\; segmentation\; task\; not\; converged\; \mathbf{do}:$ \\
        \hspace{0.25cm}\textit{[segmentor and feature extractor gradients w.r.t segmentation loss ${L_{seg}}$]}\\
        \hspace{0.25cm}$\mathbf{dW_{sh}} \leftarrow \mathbf{dW_{sh}} + \sum_{i}\delta_{i}\nabla_{W_{sh}} L(W_{sh}, W_{seg})$\\
        \hspace{0.25cm}$\mathbf{dW_{seg}} \leftarrow \mathbf{dW_{seg}} + \sum_{i}\delta_{i}\nabla_{W_{seg}} L(W_{sh}, W_{seg})$\\
        $\mathbf{end \; while}$\\
        
        \STATE \textbf{[Optimize Scene graph]}\\
        $\mathbf{while}\; scene\; graph\; task\; not\; converged\; \mathbf{do}:$ \\
        \hspace{0.25cm}\textit{[Scene graph block gradients w.r.t scene graph loss ${L_{sg}}$]}\\
        \hspace{0.25cm}$\mathbf{dW_{sg}} \leftarrow \mathbf{dW_{sg}} + \sum_{i}\delta_{i}\nabla_{W_{sg}} L(W_{sh},W_{seg},W_{sg})$\\
        $\mathbf{end \; while}$\\

    \end{algorithmic}
\end{algorithm}

\section{Experiments}
\label{Experiments}

\subsection{Dataset}
\label{dataset}
The model’s performance in interaction detection and instrument segmentation is trained and evaluated on MICCAI Endoscopic Vision Challenge 2018~\cite{allan2020018} dataset. The dataset consists of $15$ video sequences ($1$-$7$, $9$-$16$) of robotic nephrectomy procedures. 11 sequences ($2$-$4$, $6$-$7$, $9$-$12$, and $14$-$15$) are used for training and 3 sequences ($1, 5, 16$) for testing. Video sequence $13$ is not used due to a lack of tool-tissue interaction. For segmentation, eight classes: \textit{T0 - background, T1 - bipolar forceps, T2 - prograsp forceps, T3 - large needle driver, T4 - monopolar curved scissors, T5 - ultrasound probe, T6 - suction tool, and T7 - clip applier} are considered. For interaction detection, 13 types of interactions: \textit{Idle, grasping, retraction, tissue manipulation, tool manipulation, cutting, cauterization, suction, looping, suturing, clipping, staple, and ultrasound sensing} are considered~\cite{islam2020learning}.

\subsection{Implementation Details}
In our proposed model and its variants\textsuperscript{\ref{code}}, we employ cross-entropy loss to calculate the segmentation loss and multi-label loss to calculate the interaction detection loss. The models are trained using the Adam optimizer~\cite{DBLP:journals/corr/KingmaB14}. The feature extractor is initially loaded with ImageNet pre-trained weights. The learning rate at epoch = $0$ is set to $1e^{-5}$ and is decayed by 0.98 every 10 epochs. Our models are trained for 130 epochs with a batch size of 4. To reduce computational load, the input images and corresponding segmentation masks are resized from $1024$ x $1280$ to $320$ x $400$. For a fair comparison, all benchmarked SOTA models (obtained from the author's official GitHub implementations) are re-implemented and re-trained on the same dataset.

\subsection{Multi-Task Model Improving Single-Task Performance}

\begin{figure}[!b]
\centering
\includegraphics[width=0.95\linewidth]{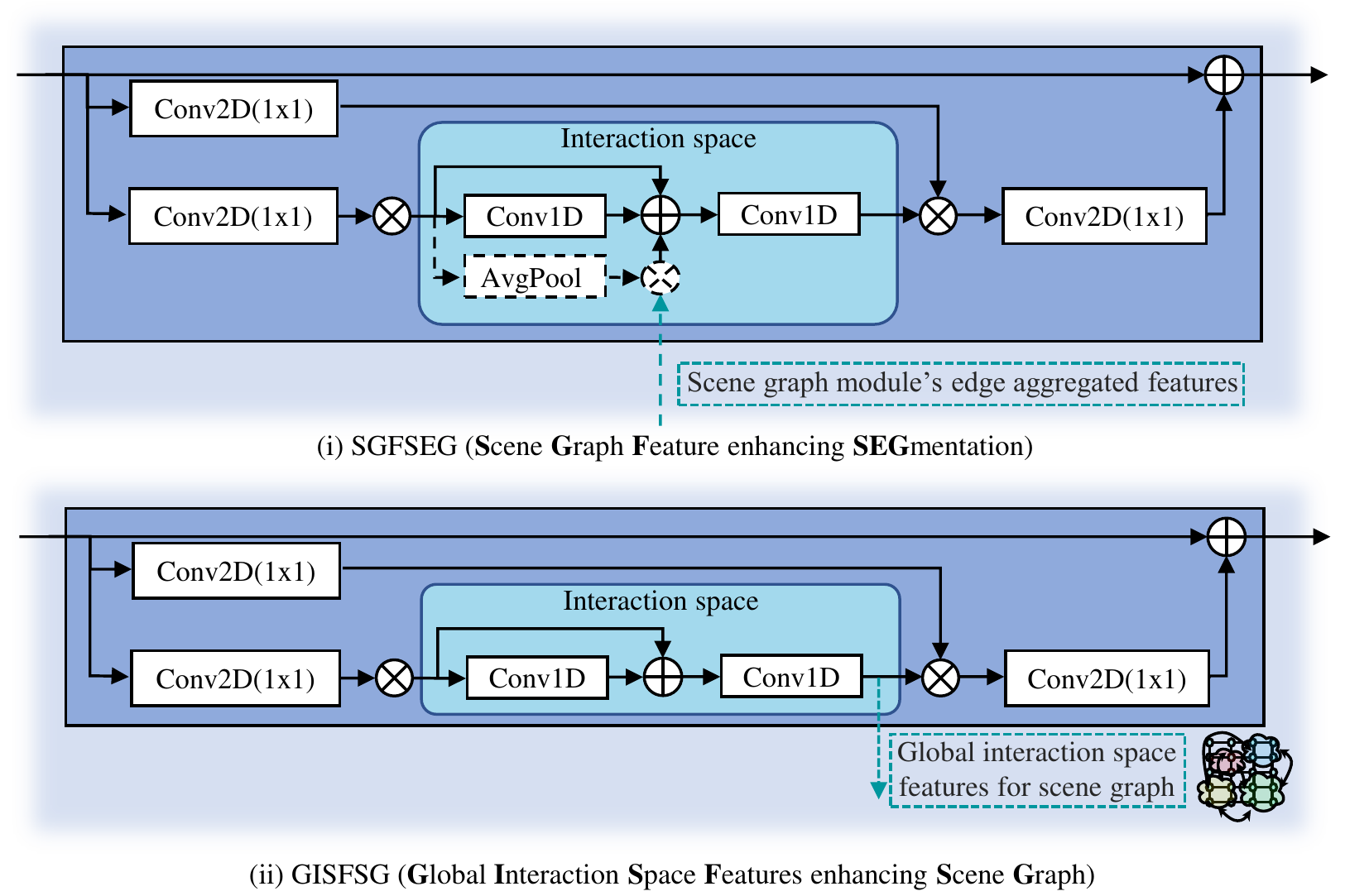}
\caption{Variants of feature sharing between the segmentation and scene graph modules in multi-task setting to improve single-task performance}
\label{fig:fig4}
\end{figure}

\begin{table*}[!t]
\centering
\caption{Comparison of our proposed globally-reasoned multi-task scene understanding model (S-MTL-MSLRGR-GISFSG) and its variant's performances against the state-of-the-art models in segmentation and tool-tissue interaction detection. T0-T7 are tool classes as stated in section \ref{dataset}.}
\scalebox{0.9}{
\begin{tabular}{cccccccccccccc}
\toprule
\multicolumn{1}{c|}{\multirow{3}{*}{\textbf{Model}}} & \multicolumn{3}{c|}{\textbf{Tool interaction detection}}                                                                                                                                                                                               & \multicolumn{8}{c}{\textbf{Segmentation}}                                                                                                                                                                                  \\ \cline{2-14} 
\multicolumn{1}{c|}{}                       & \multicolumn{1}{c|}{\multirow{2}{*}{\textbf{Acc}}} & \multicolumn{1}{c|}{\multirow{2}{*}{\textbf{mAP}}} & \multicolumn{1}{c|}{\multirow{2}{*}{\textbf{Recall}}} & \multicolumn{1}{c|}{\multirow{2}{*}{\textbf{mIoU}}} & \multicolumn{1}{c|}{\multirow{2}{*}{\textbf{P-Acc}}} & \multicolumn{8}{c}{\textbf{Class-wise IoU}}                                                                                                                                                                                       \\ \cline{7-14} 
\multicolumn{1}{c|}{}                       & \multicolumn{1}{c|}{}                     & \multicolumn{1}{c|}{}                     & \multicolumn{1}{c|}{}                        & \multicolumn{1}{c|}{}                      & \multicolumn{1}{c|}{}                      & \multicolumn{1}{c|}{T0}     & \multicolumn{1}{c|}{T1}     & \multicolumn{1}{c|}{T2}     & \multicolumn{1}{c|}{T3}     & \multicolumn{1}{c|}{T4}     & \multicolumn{1}{c|}{T5}     & \multicolumn{1}{c|}{T6}     & T7     \\ 

\toprule
\multicolumn{14}{c}{\textbf{SOTA (Surgical scene graph)}}                                                                                                                                                                                                                          \\ 
\bottomrule
\multicolumn{1}{c|}{GPNN \cite{Qi2018LearningHI}}                   & \multicolumn{1}{c|}{0.5500} & \multicolumn{1}{c|}{0.1934} & \multicolumn{1}{c|}{-}      & \multicolumn{1}{c|}{-}      & \multicolumn{1}{c|}{-}      & \multicolumn{1}{c|}{-}      & \multicolumn{1}{c|}{-}      & \multicolumn{1}{c|}{-}      & \multicolumn{1}{c|}{-}      & \multicolumn{1}{c|}{-}      & \multicolumn{1}{c|}{-}      & \multicolumn{1}{c|}{-}      & -      \\ \hline
\multicolumn{1}{c|}{Islam et al. \cite{islam2020learning}}        & \multicolumn{1}{c|}{0.4802} & \multicolumn{1}{c|}{0.2157} & \multicolumn{1}{c|}{-}      & \multicolumn{1}{c|}{-}      & \multicolumn{1}{c|}{-}      & \multicolumn{1}{c|}{-}      & \multicolumn{1}{c|}{-}      & \multicolumn{1}{c|}{-}      & \multicolumn{1}{c|}{-}      & \multicolumn{1}{c|}{-}      & \multicolumn{1}{c|}{-}      & \multicolumn{1}{c|}{-}      & -      \\ \hline
\multicolumn{1}{c|}{G-Hpooling \cite{zhang2019ierarchical}}             & \multicolumn{1}{c|}{0.3321} & \multicolumn{1}{c|}{0.1523} & \multicolumn{1}{c|}{-}      & \multicolumn{1}{c|}{-}      & \multicolumn{1}{c|}{-}      & \multicolumn{1}{c|}{-}      & \multicolumn{1}{c|}{-}      & \multicolumn{1}{c|}{-}      & \multicolumn{1}{c|}{-}      & \multicolumn{1}{c|}{-}      & \multicolumn{1}{c|}{-}      & \multicolumn{1}{c|}{-}      & -      \\ \hline
\multicolumn{1}{c|}{VS-GAT\cite{liang2020isualsemantic}}          & \multicolumn{1}{c|}{0.6537}      & \multicolumn{1}{c|}{0.2560}    & \multicolumn{1}{c|}{0.2666}      & \multicolumn{1}{c|}{-}      & \multicolumn{1}{c|}{-}      & \multicolumn{1}{c|}{-}      & \multicolumn{1}{c|}{-}      & \multicolumn{1}{c|}{-}      & \multicolumn{1}{c|}{-}      & \multicolumn{1}{c|}{-}      & \multicolumn{1}{c|}{-}      & \multicolumn{1}{c|}{-}      & -      \\ \toprule
\multicolumn{14}{c}{\textbf{SOTA (Surgical scene segmentation)}}                                                                                                                                   \\ \bottomrule
\multicolumn{1}{c|}{LinkNet34 \cite{chaurasia2017inknet}}              & \multicolumn{1}{c|}{-}      & \multicolumn{1}{c|}{-}      & \multicolumn{1}{c|}{-}      & \multicolumn{1}{c|}{0.2610}  & \multicolumn{1}{c|}{0.93}   & \multicolumn{1}{c|}{0.9193} & \multicolumn{1}{c|}{0.3581} & \multicolumn{1}{c|}{0.1481} & \multicolumn{1}{c|}{0.0062} & \multicolumn{1}{c|}{0.6488} & \multicolumn{1}{c|}{0.0004} & \multicolumn{1}{c|}{0.0071} & 0.0000 \\ \hline
\multicolumn{1}{c|}{AlbUNet \cite{shvets2018automatic}}                & \multicolumn{1}{c|}{-}      & \multicolumn{1}{c|}{-}      & \multicolumn{1}{c|}{-}      & \multicolumn{1}{c|}{0.2471}  & \multicolumn{1}{c|}{0.91}   & \multicolumn{1}{c|}{0.9090} & \multicolumn{1}{c|}{0.3610} & \multicolumn{1}{c|}{0.0923} & \multicolumn{1}{c|}{0.0064} & \multicolumn{1}{c|}{0.6082} & \multicolumn{1}{c|}{0.0000} & \multicolumn{1}{c|}{0.0000} & 0.0000 \\ \hline
\multicolumn{1}{c|}{Ternaus-UNet11 \cite{shvets2018automatic}}         & \multicolumn{1}{c|}{-}      & \multicolumn{1}{c|}{-}      & \multicolumn{1}{c|}{-}      & \multicolumn{1}{c|}{0.2406}  & \multicolumn{1}{c|}{0.917}  & \multicolumn{1}{c|}{0.8904} & \multicolumn{1}{c|}{0.3267} & \multicolumn{1}{c|}{0.0741} & \multicolumn{1}{c|}{0.0055} & \multicolumn{1}{c|}{0.6283} & \multicolumn{1}{c|}{0.0000} & \multicolumn{1}{c|}{0.0000} & 0.0000 \\ \hline
\multicolumn{1}{c|}{Ternaus-UNet16 \cite{shvets2018automatic}}         & \multicolumn{1}{c|}{-}      & \multicolumn{1}{c|}{-}      & \multicolumn{1}{c|}{-}      & \multicolumn{1}{c|}{0.2329}  & \multicolumn{1}{c|}{0.918}  & \multicolumn{1}{c|}{0.8811} & \multicolumn{1}{c|}{0.3069} & \multicolumn{1}{c|}{0.0923} & \multicolumn{1}{c|}{0.0062} & \multicolumn{1}{c|}{0.5763} & \multicolumn{1}{c|}{0.0000} & \multicolumn{1}{c|}{0.0003} & 0.0000 \\ \hline
\multicolumn{1}{c|}{MF-TAPNet \cite{jin2019ncorporating}}              & \multicolumn{1}{c|}{-}      & \multicolumn{1}{c|}{-}      & \multicolumn{1}{c|}{-}      & \multicolumn{1}{c|}{0.2489}  & \multicolumn{1}{c|}{0.931}  & \multicolumn{1}{c|}{0.9310} & \multicolumn{1}{c|}{0.2961} & \multicolumn{1}{c|}{0.0225} & \multicolumn{1}{c|}{0.0000} & \multicolumn{1}{c|}{0.7420} & \multicolumn{1}{c|}{0.0000} & \multicolumn{1}{c|}{0.0000} & 0.0000 \\ \hline
\multicolumn{1}{c|}{MF-TAPNet11 \cite{jin2019ncorporating}}            & \multicolumn{1}{c|}{-}      & \multicolumn{1}{c|}{-}      & \multicolumn{1}{c|}{-}      & \multicolumn{1}{c|}{0.3568}  & \multicolumn{1}{c|}{0.955}  & \multicolumn{1}{c|}{0.9729} & \multicolumn{1}{c|}{0.6142} & \multicolumn{1}{c|}{0.2338} & \multicolumn{1}{c|}{0.0100} & \multicolumn{1}{c|}{0.8420} & \multicolumn{1}{c|}{0.0030} & \multicolumn{1}{c|}{0.1634} & 0.0153 \\ \hline
\multicolumn{1}{c|}{MF-TAPNet34 \cite{jin2019ncorporating}}            & \multicolumn{1}{c|}{-}      & \multicolumn{1}{c|}{-}      & \multicolumn{1}{c|}{-}      & \multicolumn{1}{c|}{0.3543}  & \multicolumn{1}{c|}{0.952}  & \multicolumn{1}{c|}{\textbf{0.9767}} & \multicolumn{1}{c|}{0.6636} & \multicolumn{1}{c|}{0.3435} & \multicolumn{1}{c|}{\textbf{0.0284}} & \multicolumn{1}{c|}{0.8222} & \multicolumn{1}{c|}{0.0000} & \multicolumn{1}{c|}{0.0000} & 0.0000 \\ \hline
\multicolumn{1}{c|}{ResNet18 \cite{7780459}}                & \multicolumn{1}{c|}{-}      & \multicolumn{1}{c|}{-}      & \multicolumn{1}{c|}{-}      & \multicolumn{1}{c|}{0.3858} & \multicolumn{1}{c|}{0.9487} & \multicolumn{1}{c|}{0.9533} & \multicolumn{1}{c|}{0.5764} & \multicolumn{1}{c|}{0.3810} & \multicolumn{1}{c|}{0.0008} & \multicolumn{1}{c|}{0.8353} & \multicolumn{1}{c|}{0.0073} & \multicolumn{1}{c|}{0.2763} & \textbf{0.0557} \\ \hline
\multicolumn{1}{c|}{ResNet18 + GloRe~\cite{chen2019graph}}        & \multicolumn{1}{c|}{-}      & \multicolumn{1}{c|}{-}      & \multicolumn{1}{c|}{-}      & \multicolumn{1}{c|}{0.3926} & \multicolumn{1}{c|}{0.9483} & \multicolumn{1}{c|}{0.9524} & \multicolumn{1}{c|}{0.5764} & \multicolumn{1}{c|}{0.3842} & \multicolumn{1}{c|}{0.0009} & \multicolumn{1}{c|}{0.8256} & \multicolumn{1}{c|}{0.0720} & \multicolumn{1}{c|}{0.2847} & 0.0448 \\
\toprule
%\multicolumn{14}{c}{Ours}                                                                                                                                                                                                                                                                                                                                                                                                    \\ 
%\toprule
\multicolumn{14}{c}{\textbf{S-MTL (Ours)}}                                                                                                                                                                                                                                                                                                                                                                                             \\ 
\bottomrule

\multicolumn{1}{c|}{MSLRGR}            & \multicolumn{1}{c|}{\textbf{0.7003}} & \multicolumn{1}{c|}{0.2885} & \multicolumn{1}{c|}{0.3096} & \multicolumn{1}{c|}{\textbf{0.4354}} & \multicolumn{1}{c|}{\textbf{0.9638}} & \multicolumn{1}{c|}{0.9714} & \multicolumn{1}{c|}{\textbf{0.6966}} & \multicolumn{1}{c|}{\textbf{0.4356}} & \multicolumn{1}{c|}{0.0015} & \multicolumn{1}{c|}{\textbf{0.8716}} & \multicolumn{1}{c|}{0.1203} & \multicolumn{1}{c|}{0.3471} & 0.0387 \\ \hline
\multicolumn{1}{c|}{MSLRGR-GISFSG}            & \multicolumn{1}{c|}{0.6994} & \multicolumn{1}{c|}{\textbf{0.3131}} & \multicolumn{1}{c|}{\textbf{0.3157}} & \multicolumn{1}{c|}{\textbf{0.4354}} & \multicolumn{1}{c|}{\textbf{0.9638}} & \multicolumn{1}{c|}{0.9714} & \multicolumn{1}{c|}{\textbf{0.6966}} & \multicolumn{1}{c|}{\textbf{0.4356}} & \multicolumn{1}{c|}{0.0015} & \multicolumn{1}{c|}{\textbf{0.8716}} & \multicolumn{1}{c|}{0.1203} & \multicolumn{1}{c|}{0.3471} & 0.0387 \\ 
\toprule
\multicolumn{14}{c}{\textbf{KD-MTL (Ours)}}       

\\ \hline
\multicolumn{1}{c|}{MSLRGR}                     & \multicolumn{1}{c|}{0.6434}       & \multicolumn{1}{c|}{0.2992}       & \multicolumn{1}{c|}{0.2818}       & \multicolumn{1}{c|}{0.4105}       & \multicolumn{1}{c|}{0.9617}       & \multicolumn{1}{c|}{0.9704}       & \multicolumn{1}{c|}{0.6775}       & \multicolumn{1}{c|}{0.3650}       & \multicolumn{1}{c|}{0.0002}       & \multicolumn{1}{c|}{0.8598}       & \multicolumn{1}{c|}{0.0709}       & \multicolumn{1}{c|}{0.3330}       & 0.0069       \\ \hline

\multicolumn{1}{c|}{MSLRGR-GISFSG}                     & \multicolumn{1}{c|}{0.6710}       & \multicolumn{1}{c|}{0.2808}       & \multicolumn{1}{c|}{0.3072}       & \multicolumn{1}{c|}{0.4105}       & \multicolumn{1}{c|}{0.9611}       & \multicolumn{1}{c|}{0.9713}       & \multicolumn{1}{c|}{0.6670}       & \multicolumn{1}{c|}{0.3785}       & \multicolumn{1}{c|}{0.0028}       & \multicolumn{1}{c|}{0.8603}       & \multicolumn{1}{c|}{0.0458}       & \multicolumn{1}{c|}{0.3184}       &    0.0401    \\ 
\bottomrule
\label{tab-main}
\end{tabular}}

\end{table*}

\begin{figure*}[!h]
\centering
\includegraphics[width=0.9\linewidth]{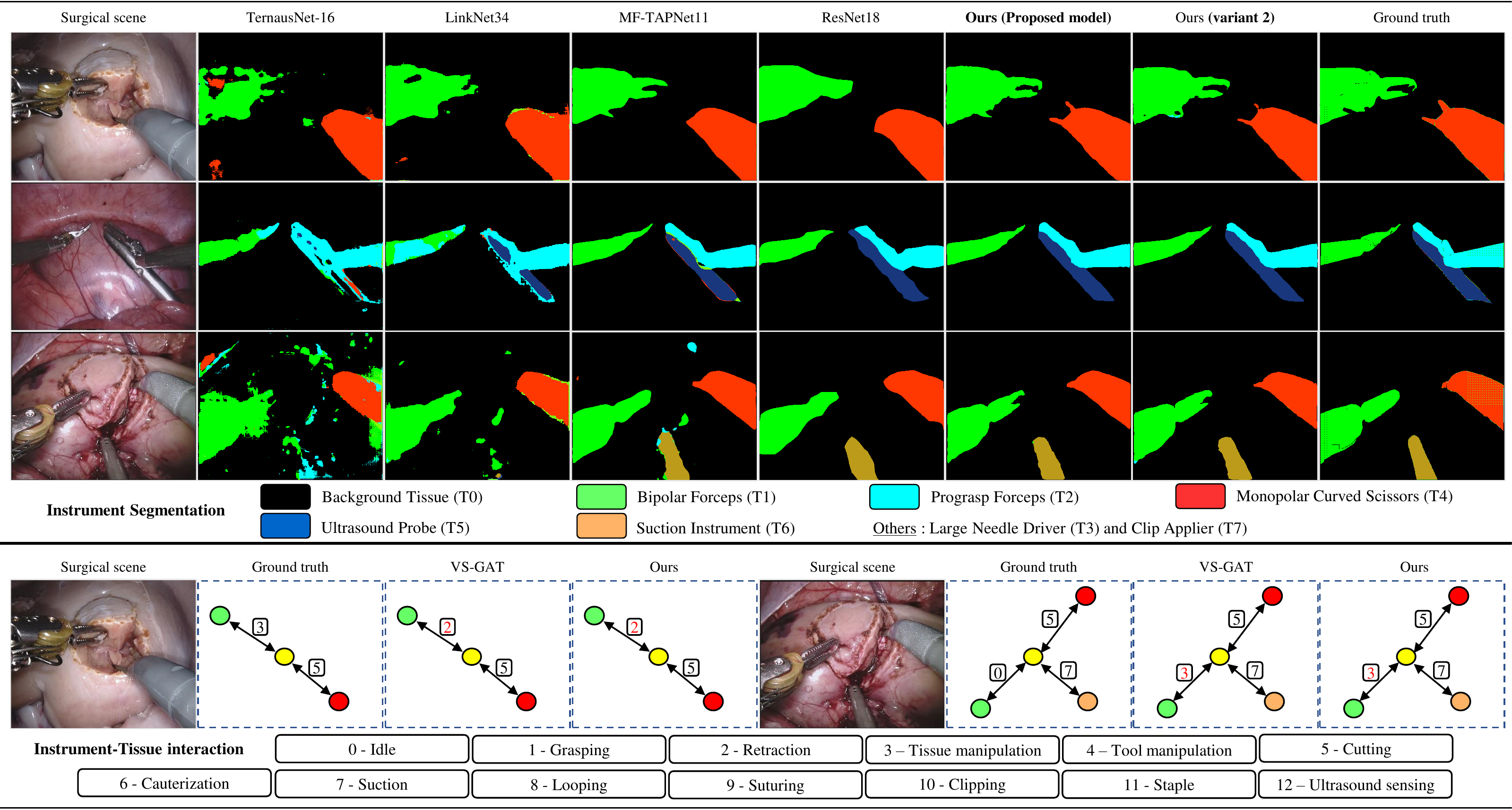}
\caption{Qualitative analysis: \textbf{Top} - Comparison of our proposed model and its variant’s performance in instrument segmentation against select benchmark models and the Ground Truth (GT). \textbf{Bottom} - Comparison of our proposed model’s performance in interaction detection against using vanilla VS-GAT~\cite{liang2020isualsemantic} and the Ground Truth (GT). Here, our proposed model refers to S-MTL-MSLRGR-GISFSG (sequentially trained multi-task learning model with multi-scale local reasoning and global reasoning and its scene graph enhanced with global interaction space features).}
\label{fig:fig5}
\end{figure*}

% Please add the following required packages to your document preamble:
% \usepackage{multirow}
\begin{table*}[h]
\centering
\caption{Ablation study highlighting the importance of multi-scale local and global reasoning (MSLRGR) and use of global interaction space feature in the scene graph (GISFSG) in improving sequentially optimized multi-task learning (S-MTL) model.}
\scalebox{0.90}{
\begin{tabular}{ccccccccccc}
\toprule
\multicolumn{1}{c|}{\multirow{2}{*}{\textbf{Model}}} & \multicolumn{3}{c|}{\textbf{Feature encoder}}                                                    & \multicolumn{2}{c|}{\textbf{SF}}                                                                 & \multicolumn{3}{c|}{\textbf{Tool interaction detection}}                                                                                         & \multicolumn{2}{c}{\textbf{Segmentation}}                      \\ \cline{2-11} 
\multicolumn{1}{c|}{}                       & 
\multicolumn{1}{c|}{\textbf{GR~\cite{chen2019graph}}}                &
\multicolumn{1}{c|}{\textbf{MSGR}}                & \multicolumn{1}{c|}{\textbf{MSLR}}                    & \multicolumn{1}{c|}{\textbf{PF}}                   & \multicolumn{1}{c|}{\textbf{GISFSG}}                 & \multicolumn{1}{c|}{\textbf{Acc}}             & \multicolumn{1}{c|}{\textbf{mAP}}             & \multicolumn{1}{c|}{\textbf{Recall}}          & \multicolumn{1}{c|}{\textbf{mIoU}}            & \multicolumn{1}{c}{\textbf{P-Acc}}           \\ \toprule
\multicolumn{11}{c}{STL}                                                                                                                                                                                                                                                                                                                                                                        \\ \bottomrule
\multicolumn{1}{c|}{VS-GAT~\cite{liang2020isualsemantic}}    & \multicolumn{1}{c|}{}          & \multicolumn{1}{c|}{}                      & \multicolumn{1}{c|}{}                      & \multicolumn{1}{c|}{}                      & \multicolumn{1}{c|}{}                      & \multicolumn{1}{c|}{0.6537}          & \multicolumn{1}{c|}{0.2560}           & \multicolumn{1}{c|}{0.2666}          & \multicolumn{1}{c|}{-}                & -                 \\ \hline
\multicolumn{1}{c|}{SEG}  
& \multicolumn{1}{c|}{} & \multicolumn{1}{c|}{}                      & \multicolumn{1}{c|}{}                      & \multicolumn{1}{c|}{}                      & \multicolumn{1}{c|}{}                      & \multicolumn{1}{c|}{-}                & \multicolumn{1}{c|}{-}                & \multicolumn{1}{c|}{-}                & \multicolumn{1}{c|}{0.3858}          & 0.9487          \\ \hline
\multicolumn{1}{c|}{SEG-GR}  
& \multicolumn{1}{c|}{\cmark} & \multicolumn{1}{c|}{}                      & \multicolumn{1}{c|}{}                      & \multicolumn{1}{c|}{}                      & \multicolumn{1}{c|}{}                      & \multicolumn{1}{c|}{-}                & \multicolumn{1}{c|}{-}                & \multicolumn{1}{c|}{-}                & \multicolumn{1}{c|}{0.3926}          & 0.9483          \\ \hline
\multicolumn{1}{c|}{SEG-MSGR} 
& \multicolumn{1}{c|}{\cmark} & \multicolumn{1}{c|}{\cmark} & \multicolumn{1}{c|}{}                      & \multicolumn{1}{c|}{}                      & \multicolumn{1}{c|}{}                      & \multicolumn{1}{c|}{-}                & \multicolumn{1}{c|}{-}                & \multicolumn{1}{c|}{-}                & \multicolumn{1}{c|}{0.4350}           & 0.9628          \\ \hline
\multicolumn{1}{c|}{SEG-MSLRGR}  
& \multicolumn{1}{c|}{\cmark} & \multicolumn{1}{c|}{}                      & \multicolumn{1}{c|}{\cmark} & \multicolumn{1}{c|}{}                      & \multicolumn{1}{c|}{}                      & \multicolumn{1}{c|}{-}                & \multicolumn{1}{c|}{-}                & \multicolumn{1}{c|}{-}                & \multicolumn{1}{c|}{\textbf{0.4354}} & \textbf{0.9638} \\ \toprule
\multicolumn{11}{c}{S-MTL}                                                                   \\ \bottomrule
\multicolumn{1}{c|}{GR} 
& \multicolumn{1}{c|}{\cmark} & \multicolumn{1}{c|}{}                      & \multicolumn{1}{c|}{}                      & \multicolumn{1}{c|}{}                      & \multicolumn{1}{c|}{}                      & \multicolumn{1}{c|}{0.6787}          & \multicolumn{1}{c|}{0.2578}          & \multicolumn{1}{c|}{0.3042}          & \multicolumn{1}{c|}{0.3926}          & 0.9483          \\ \hline
\multicolumn{1}{c|}{MSGR} 
& \multicolumn{1}{c|}{\cmark} & \multicolumn{1}{c|}{\cmark} & \multicolumn{1}{c|}{}                      & \multicolumn{1}{c|}{}                      & \multicolumn{1}{c|}{}                      & \multicolumn{1}{c|}{0.6813}          & \multicolumn{1}{c|}{0.2906}          & \multicolumn{1}{c|}{0.3040}          & \multicolumn{1}{c|}{0.4350}          & 0.9628          \\ \hline
\multicolumn{1}{c|}{MSLRGR}  
& \multicolumn{1}{c|}{\cmark} & \multicolumn{1}{c|}{}                      & \multicolumn{1}{c|}{\cmark} & \multicolumn{1}{c|}{}                      & \multicolumn{1}{c|}{}                      & \multicolumn{1}{c|}{\textbf{0.7003}} & \multicolumn{1}{c|}{0.2885}          & \multicolumn{1}{c|}{0.3096}          & \multicolumn{1}{c|}{0.4354}          & 0.9638          \\ \hline
\multicolumn{1}{c|}{MSLRGR-PF}  
& \multicolumn{1}{c|}{\cmark} & \multicolumn{1}{c|}{}                      & \multicolumn{1}{c|}{\cmark} & \multicolumn{1}{c|}{\cmark} & \multicolumn{1}{c|}{}                      & \multicolumn{1}{c|}{0.6848}          & \multicolumn{1}{c|}{0.2960}           & \multicolumn{1}{c|}{\textbf{0.3157}} & \multicolumn{1}{c|}{\textbf{0.4354}} & \textbf{0.9638} \\ \hline
\multicolumn{1}{c|}{\textbf{MSLRGR-GISFSG}}   
& \multicolumn{1}{c|}{\cmark} & \multicolumn{1}{c|}{}                      & \multicolumn{1}{c|}{\cmark} & \multicolumn{1}{c|}{}                      & \multicolumn{1}{c|}{\cmark} & \multicolumn{1}{c|}{0.6994}          & \multicolumn{1}{c|}{\textbf{0.3131}} & \multicolumn{1}{c|}{\textbf{0.3157}} & \multicolumn{1}{c|}{\textbf{0.4354}} & \textbf{0.9638} \\ 

\bottomrule
\label{table_s_mtl}
\end{tabular}}
\end{table*}

\begin{table*}[!h]
    \centering
    \caption{Ablation Study on multi-task learning (MTL) model optimized using Vanilla-MTL (V-MTL) and Knowledge Distillation-based MTL (KD-MTL) optimization techniques.}
    \scalebox{0.95}{
    \begin{tabular}{c|c|c|c|c|c|c|c|c|c|c|c|c}

    \toprule

    \multirow{4}{*}{\textbf{Model}} & \multicolumn{4}{c|}{\textbf{Best in tool interaction detection}}                                                  & \multicolumn{4}{c|}{\textbf{Best in instrument segmentation}}                                                     & \multicolumn{4}{c}{\textbf{Balanced performance}}                                                                 \\ \cline{2-13} 
                                    & \multicolumn{2}{c|}{\textbf{Tool interaction}}    & \multicolumn{2}{c|}{\multirow{2}{*}{\textbf{Segmentation}}}   & \multicolumn{2}{c|}{\textbf{Tool interaction}}    & \multicolumn{2}{c|}{\multirow{2}{*}{\textbf{Segmentation}}}   & \multicolumn{2}{c|}{\textbf{Tool interaction}}    & \multicolumn{2}{c}{\multirow{2}{*}{\textbf{Segmentation}}}    \\
                                    & \multicolumn{2}{c|}{\textbf{detection}}           & \multicolumn{2}{c|}{}                                       & \multicolumn{2}{c|}{\textbf{detection}}          & \multicolumn{2}{c|}{}                                        & \multicolumn{2}{c|}{\textbf{detection}}           & \multicolumn{2}{c}{}                                          \\ \cline{2-13} 
                                    & \textbf{Acc}     & \textbf{mAP}                   & \textbf{mIoU}        & \textbf{P-Acc}                         & \textbf{Acc}     & \textbf{mAP}                   & \textbf{mIoU}        & \textbf{P-Acc}                         & \textbf{Acc}     & \textbf{mAP}                   & \textbf{mIoU}        & \textbf{P-Acc}                         \\ \hline
    V-MTL-GR                        & 0.6193           & 0.2303                         & 0.3521               & 0.9420                                  & 0.5073           & 0.2580                          & 0.3731               & 0.9455                                 & 0.6064           & 0.2327                         & 0.3621               & 0.9447                                 \\ \hline
    KD-MTL-GR                       & 0.6615           & 0.2531                         & 0.3609               & 0.9449                                 & \textbf{0.6391}  & 0.2472                         & 0.3730                & 0.9453                                 & 0.6555           & 0.2522                         & 0.3713               & 0.9458                                 \\ \hline
    KD-MTL-MSLRGR                   & 0.6649           & 0.2644                         & 0.4022               & 0.9610                                  & 0.6322           & 0.2724                         & 0.4165               & 0.9622                                 & 0.6434           & \textbf{0.2992}                & 0.4105               & \textbf{0.9617}                        \\ \hline
    KD-MTL-MSLRGR-SGFSEG            & 0.6589           & 0.2636                         & 0.3974               & 0.9593                                 & 0.6184           & \textbf{0.2829}                & \textbf{0.4188}      & 0.9607                                 & 0.6503           & 0.2600                         & \textbf{0.4111}      & 0.9608                                 \\ \hline
    KD-MTL-MSLRGR-GISFSG            & \textbf{0.6830}  & \textbf{0.2818}                & \textbf{0.4034}      & \textbf{0.9613}                        & 0.6339           & 0.2819                         & 0.4169               & \textbf{0.9617}                        & \textbf{0.6710}  & 0.2808                         & 0.4105               & 0.9611                                 \\
    \bottomrule
    \end{tabular}}
    \label{tab_mtl_ablation}
\end{table*}

MTL models are used to reduce the computational cost by sharing common modules between the two tasks. Additionally, we experiment to improve its performance over the STL models through the multi-task model setup. In achieving this, three variants of MTL models have been explored. Firstly, in the Vanilla-MTL model, the two task models are naively implemented with only its feature encoder shared. Here, we aim to improve the interaction detection model. Since the shared feature encoder is used to extract features for segmentation, the encoder is better trained to encode tool features, improving interaction detection. The second variant (Fig. \ref{fig:fig4} (i)) aims at utilizing scene graph features to enhance segmentation (SGFSEG). In this variant, the interaction features from the VS-GAT’s combined graph ($\mathcal{G}_C$) edges are appended to the latent interaction space features in the segmentation module’s GloRe unit. In the final variant (used in our final proposed model), in addition to the advantage of vanilla MTL, we aim to use global interaction space features to improve scene graph (GISFSG) interaction detection (Fig. \ref{fig:fig4} (ii)). The globally-reasoned feature from the latent interaction space in the GloRe unit is appended to the edges of the combined interaction detection graph ($\mathcal{G}_C$). This allows the graph network to detect interaction based on globally-reasoned node-to-node interaction features.

\subsection{Results and Evaluation}
The performance of our proposed global-reasoned multi-task scene understanding model in segmenting instruments and detecting interaction is analyzed both quantitatively and qualitatively. Quantitatively, the model's performance in both tasks is benchmarked against its respective single task SOTA models. The performance in instrument segmentation is quantified using (a) the mean intersection over onion (mIoU), class-wise IoU, and pixel accuracy (P-Acc) metrics. The performance in interaction detection is quantified using accuracy (Acc), mean average precision (mAP), and Recall. From table \ref{tab-main}, it is observed that our globally-reasoned multi-task model (S-MTL-MSLRGR-GISFSG) performance is on par and, in most cases, outperforms STL models in both instrument segmentation (mIoU and P-Acc) and interaction detection (Acc, mAP, and Recall). This observation is further validated based on qualitative performance (Fig. \ref{fig:fig5}). Qualitatively, it is also observed that the model's performance with global reasoning in latent space is further enhanced by incorporating multi-scale local reasoning. Here, the segmentation performance of the SOTA models are significantly different from their original works due to three main reasons: (i) Change in train and test set, (ii) change in number and type of classes and (iii) change in resolution of the input image.

\subsection{Ablation Studies}
We report a detailed ablation study in Table \ref{table_s_mtl} that demonstrates the significance of (i) MSLRGR in improving instrument segmentation performance, (ii) multi-task model setup and, (iii) embedding scene graph module with globally-reasoned latent features in improving interaction detection. Firstly, improvement in segmentation due to MSLRGR (mIoU = $0.4354$) against just using GR (mIoU = $0.3926$) and MSGR (mIoU = $0.4350$) is reported in both single-task and sequentially optimized multi-task model setup. This proves that incorporating global latent interaction reasoning and multi-scale local (neighborhood) reasoning improves segmentation. Secondly, it is also observed that the STL interaction model (VS-GAT~\cite{liang2020isualsemantic}) is outperformed by the multi-task model (S-MTL-GR). This performance is attributed to better feature embedding by the feature encoder module also trained for instrument segmentation. Thirdly, we also observe that when a globally-reasoned latent interaction feature is incorporated into the scene graph model (S-MTL-MSLRGR-GISFSG), its interaction detection performance is improved (mAP = $0.3131$). Comparing this performance against the performance (mAP = $0.296$) of the scene graph model embedded with penultimate features (S-MTL-MSLRGR-PF) further proves that globally-reasoned interaction space features significantly improves interaction detection. 

On top of reporting an ablation study on the model’s performance enhancement from global reasoning, an ablation study on the model’s performance based on multi-task model optimization has also been reported. From Table \ref{tab_mtl_ablation}, it is observed that KD-MTL optimization (KD-MTL-GR) outperforms the Vanilla-MTL optimization technique (V-MTL-GR).  In both these techniques, it is observed that achieving a synchronous convergence in its task performances is still a problem, making the sequential optimization technique (reported in Table \ref{table_s_mtl}) superior. Based on KD-MTL optimization performance, it is observed that segmentation performance is slightly improved (KD-MTL-MSLRGR-SGFSEG: mIoU = $0.4111$) if the scene graph module’s interaction features are embedded into the GR unit’s latent interaction space. However, this is still outperformed by S-MTL models.

The results in Table.~\ref{tab-main}, Table.~\ref{table_s_mtl} and Table.~\ref{tab_mtl_ablation} are obtained based on a single-fold validation approach as the dataset is restricted by long-tailed class distribution. The instrument class distribution across the sequences is severely imbalanced, with some instruments (ultrasound probe, large needle driver, suturing, and clip applier) appearing only in one or two sequences. Therefore, sequences 1, 5, and 16 were carefully chosen as the test set to ensure that optimal instrument class distribution is achieved across the train and test set. However, to validate our model’s superior performance, key models are trained and tested using the 4-fold (Test sequences = \{1,5,16\}, \{2,3,15\}, \{4,6,14\} and \{4,11,12\}) cross-validation approach. From Table.~\ref{table:multi_fold_performance}, it is observed that our model (S-MTL-MSLRGR-GISFSG) outperforms other key models based on the average performance in multi-fold cross-validation approach.

\begin{table}[!h]
    
    \centering
    \caption{Multi-fold cross-validation}
    \scalebox{0.9}{
    \begin{tabular}{c|c|c|c|c|c}
    \toprule
    \multirow{2}{*}{\textbf{Model}}                      & \multicolumn{3}{c|}{\textbf{Tool interaction detection}}                                                  & \multicolumn{2}{c}{\textbf{Segmentation}}                             \\ \cline{2-6}
                                                         & \textbf{Acc}                      & \textbf{mAP}                      & \textbf{Recall}                   & \textbf{mIoU}                     & \textbf{P-Acc}                    \\
    \midrule   
    MF-TAPNet11~\cite{jin2019ncorporating}               & -                                 & -                                 & -                                 & 0.2659                            & 0.9330                            \\ \hline
    ResNet18\cite{7780459} + GloRe~\cite{chen2019graph}  & -                                 & -                                 & -                                 & 0.3624                            & 0.9209                            \\ \hline
    VS-GAT~\cite{liang2020isualsemantic}                 & 0.5911                            & 0.2436                            & 0.2805                            & -                                 & -                                 \\ \hline
    S-MTL (Ours)                                         & \multirow{2}{*}{\textbf{0.6168}}  & \multirow{2}{*}{\textbf{0.2662}}  & \multirow{2}{*}{\textbf{0.2917}}  & \multirow{2}{*}{\textbf{0.3786}}  & \multirow{2}{*}{\textbf{0.9341}}  \\
    MSLRGR-GISFSG                                        &                                   &                                   &                                   &                                   &                                   \\
    \bottomrule
    \end{tabular}}
    \label{table:multi_fold_performance}
    
\end{table}

\begin{table}[!b]
    
    \centering
    \caption{Computation performance}
    \scalebox{0.85}{
    \begin{tabular}{c|c|c|c|c}
    \toprule
    \multirow{2}{*}{\textbf{Model}}                      & \textbf{Model}           & \textbf{Parameter}         & \textbf{Training}       & \textbf{Inference}        \\
                                                         & \textbf{size (MB)}       & \textbf{count}             & \textbf{time (hrs)}     & \textbf{time (ms)}        \\
    \midrule
    ResNet18\cite{7780459} + GloRe~\cite{chen2019graph}  & 53.70                    & 14041448                   & 5.92                    & 51.3885                   \\ \hline
    VS-GAT~\cite{liang2020isualsemantic}                 & 51.88                    & 13574307                   & 4.90                    & 52.4388                   \\ \hline
    S-MTL (Ours)                                         & \multirow{2}{*}{63.346}  & \multirow{2}{*}{16566211}  & \multirow{2}{*}{10.89}  & \multirow{2}{*}{54.7293}  \\
    MSLRGR-GISFSG                                        &                          &                            &                         &                           \\ \hline
    KD-MTL (Ours)                                        & \multirow{2}{*}{63.346}  & \multirow{2}{*}{16566211}  & \multirow{2}{*}{17.58}  & \multirow{2}{*}{54.5400}  \\
    MSLRGR-GISFSG                                        &                          &                            &                         &                           \\
    \bottomrule
    \end{tabular}
    \label{table:computation_performance}
    }
    
\end{table}

Finally, our proposed model is also evaluated based on computational performance (Table.~\ref{table:computation_performance}). While the S-MTL model takes around $10.89$ hrs to train, the KD-MTL model takes $17.58$ hrs to train. The KD-MTL is reported to have high training time as we take into account the time ($10.82$ hrs) taken to train the two independent teacher models: (i) segmentation (ResNet18~\cite{7780459} + GloRe~\cite{chen2019graph}) and (ii) tool interaction detection model (VS-GAT\cite{liang2020isualsemantic}). If we disregard the training time of the teacher models in KD-MTL, the resultant training time is $6.76$ hrs. However, in this case, the time taken to train the segmentation model in the S-MTL ($5.92$ hrs) must be subtracted for a fair comparison. In the S-MTL training regime, stage $3$ (\emph{Optimize feature extractor and segmentation network} in Algorithm~\ref{algortihm_ato}) can be skipped by loading the pre-trained weights from the segmentation model resulting in a reduced training time of $4.97$ hrs. Thus, using the S-MTL model over the KD-MTL model is also justified based on computational performance. Additionally, it is also observed that the size of our MTL model is significantly smaller than the size of the two independent models combined. In terms of inference time, both S-MTL and KD-MTL takes $\sim 54ms$ for each frame which is suitable for real-time application.

\section{Discussion and Conclusion}
\label{Discussion and Conclusion}
 In the paper, a globally-reasoned multi-task surgical scene understanding model to perform instrument segmentation and tool-tissue interaction detection is proposed. The model’s performance is improved by (i) introducing multi-scale local (neighborhood) reasoning and incorporating latent global reasoning and (ii) introducing global interaction space features into the scene graph.  The detailed study also proves that the proposed model performs on-par and, in most cases, outperforms existing SOTA single-task models in MICCAI endoscopic vision challenge 2018. We also report the performances of sequential and distillation-based MTL optimization techniques and conclude that sequential training results in optimal convergence. While improvement in segmentation task from incorporating it with scene graph is less significant, it is mainly attributed to asynchronous converge of the model. Additionally, we further prove our model superiority using a multi-fold cross-validation approach. Computationally, our MTL model is also smaller than the two independent models combined and can be used for real-time applications. Further research is needed on synchronous optimization of the MTL model for the sub-tasks to benefit mutually.

\section*{ACKNOWLEDGMENT}
\label{acknowledgement}
This work was supported by the National Key R\&D Program of China under Grant 2018YFB1307700 (with subprogram 2018YFB1307703) from the Ministry of Science and Technology (MOST) of China, Hong Kong Research Grants Council (RGC) Collaborative Research Fund (CRF C4026-21GF), the Shun Hing Institute of Advanced Engineering (SHIAE project BME-p1-21, 8115064) at the Chinese University of Hong Kong (CUHK), and Singapore Academic Research Fund under Grant R397000353114.

%%%%%%%%%%%%%%%%%%%%%%%%%%%%%%%%%%%%%%%%%%%%%%%%%%%%%%%%%%%%%%%%%%%%%%%%%%%%%%%%
\balance
\bibliographystyle{IEEEtran}
\bibliography{sample}

\end{document}